\documentclass[aip, apl,reprint, floatfix]{revtex4-1}
\usepackage{amsfonts, amssymb, amsbsy, amsmath}
\usepackage{graphicx}

\begin{document}

\title{Non-hysteretic superconducting quantum interference proximity transistor with enhanced responsivity }

\newcommand{\figta}{$\left(\mathrm{a}\right)\;$}
\newcommand{\figtb}{$\left(\mathrm{b}\right)\;$}
\newcommand{\figtc}{$\left(\mathrm{c}\right)\;$}
\newcommand{\figtd}{$\left(\mathrm{d}\right)\;$}
\newcommand{\figte}{$\left(\mathrm{e}\right)\;$}
\newcommand{\figtf}{$\left(\mathrm{f}\right)\;$}
\newcommand{\figa}{$\left(\mathrm{a}\right)$}
\newcommand{\figb}{$\left(\mathrm{b}\right)$}
\newcommand{\figc}{$\left(\mathrm{c}\right)$}
\newcommand{\figd}{$\left(\mathrm{d}\right)$}
\newcommand{\fige}{$\left(\mathrm{e}\right)$}
\newcommand{\figf}{$\left(\mathrm{f}\right)$}

\thispagestyle{empty}
\setcounter{page}{1}
\clearpage
\author{R. N. Jabdaraghi}
\email{robab.najafi.jabdaraghi@aalto.fi}
\affiliation{O.V. Lounasmaa Laboratory, Aalto University School of Science, POB 13500, FI-00076
AALTO, Finland}
 \affiliation{Faculty of physics, University of Tabriz, Tabriz, Iran}

\author{M. Meschke}
\affiliation{O.V. Lounasmaa Laboratory, Aalto University School of Science, POB 13500, FI-00076
AALTO, Finland} 

\author{J. P. Pekola}
\affiliation{O.V. Lounasmaa Laboratory, Aalto University School of Science, POB 13500, FI-00076
AALTO, Finland} 

\begin{abstract}
This letter presents fabrication and characterization of an optimized SQUIPT (superconducting quantum interference proximity transistor). The present device, characterized with reduced tunnel junction area and shortened normal-metal section, demonstrates no hysteresis at low temperatures as we increased the Josephson inductance of the weak link by decreasing its cross section. It has consequently almost an order of magnitude improved magnetic field responsivity as compared to the earlier design. The modulation of both the current and the voltage across the junction have been measured as a function of magnetic flux piercing the superconducting loop. 
\end{abstract}

\date{\today}

\maketitle

An interferometer based on the proximity effect,~\cite{Gennes66,Tinkham96,Gennes64,Buzdin05,Belzing99,Pannetier0,Zhou98} the superconducting quantum interference proximity transistor (SQUIPT),~\cite{Giazotto10,Meschke11} has been introduced to detect small magnetic fields and moments. A specific application of the SQUIPT as a sensitive magnetometer has been reported with attractive features including ultra-low dissipation, a simple DC read out scheme, and flexibility in fabrication materials and parameters. Similar to nanoSQUIDs, there is a large number of foreseen applications for SQUIPT devices including measurement of magnetic flux induced by atomic spins, single-photon detection and nanoelectronical measurements.~\cite{Foley09,Hao05,Hao07} The SQUIPT has been discussed as a flux-to-voltage and flux-to-current transformer both theoretically and experimentally.~\cite{Giazotto10,Meschke11} Andreev ''mirrors''~\cite{Andreev 64} with varying phase differences between superconductors provide a means to modulate the supercurrent through a short normal wire as was shown in experiments by Petrashov $et~ al$.~\cite{Petrashov94,Petrashov95,Belzing02} Recently, incorporation of ferromagnetic layers into the SQUIPT design ~\cite{Alidoust13} and both hysteretic and non-hysteretic behavior of Andreev interferometers with three superconducting electrodes in voltage biased ~\cite{Galaktionov12} and current biased regime ~\cite {Galaktionov13} have been theoretically investigated. 

A SQUIPT consists of a superconducting loop interrupted by a normal-metal island in clean contact with it while an Al probe is tunnel-coupled to the middle of the normal region. A detailed view of this device is shown in Fig. 1 where the geometry of the island is determined by weak-link width $d$, length $L$ and its thickness $a$.
 A recent work~\cite{Meschke11} reported an advanced version of SQUIPT characterized by $d=200$ nm, $L=300$ nm and $a=20$ nm respectively, reaching current responsivity on magnetic flux of 3 nA$/ \Phi_0$. Here, $ \Phi_0=h/(2e)$ is the superconducting flux quantum. However, the usability of these devices was limited by the hysteresis appearing at low temperatures.

\begin{figure}[htb] 
\includegraphics[width = 1\columnwidth]{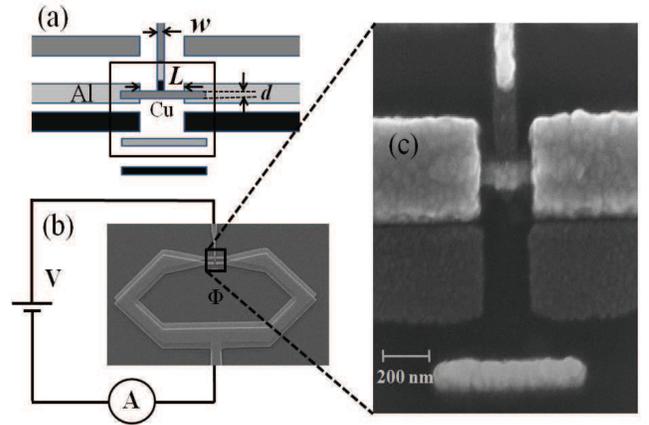}
\caption{\figta  Schematic view of our design and shadow mask layout. The three angle evaporations, consisting of Al at $31^0$ (black), Cu at $-31^0$  (dark grey), Al at $-10^0$  (light grey). The black box marks the area of the SEM image that is depicted in (c).  (b) Sketch of the electrical setup of a SQUIPT attached to a voltage bias and current measurement, $\Phi$ represents the external magnetic flux through the Al superconducting loop. (c) SEM micro-graph of the sample core illustrating Cu island embedded between two Al superconductors. The Al probe connected to the middle of the normal-metal forms the normal metal-insulator-superconductor (NIS) junction.}

\label{fig:sample}
\end{figure}
 Hysteresis in this device appears when the self-inductance of the superconducting ring well exceeds the Josephson inductance $L_J=\Phi_0/(2 \pi I_C)$, where $I_C$ is the critical current of the superconductor-normal-metal-superconductor (SNS) junction.~\cite{You11 }
For the ideal junction, $I_C\propto R_N^{-1}$ in which $R_N=\rho l/A$  is the normal-state resistance of the SNS junction, $\rho=1/ (\nu_F e^2 D)$ is the island resistivity, $A$ is the island cross section, $\nu_F$  is the density of states at the Fermi level in N and $D$ is the diffusion coefficient of the normal metal.~\cite{Giazotto08,Blum04,Dubos08}
The hysteresis is suppressed at low temperatures by increasing the Josephson inductance of the junction.~\cite{Dubos08,Heikkila02} We achieved this by shrinking the cross section $A$ of the weak-link which leads to an increased normal-state resistance and Josephson inductance.
\begin{figure}[htb] 
\includegraphics[width = 1.0\columnwidth]{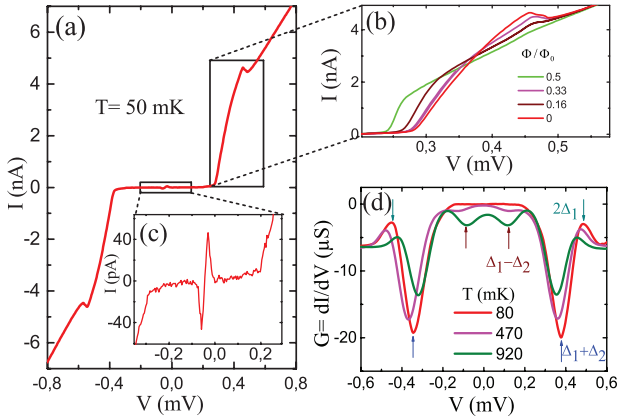}
\caption{\figta   I-V characteristics of sample A measured at $ T_{bath}=50$ mK. (b) An enlarged view of current-voltage curve at several values of magnetic flux $\Phi$  between 0 and $0.5\Phi_0$. (c) The magnitude of supercurrent appearing around zero bias voltage in device A is $I_S= 48$ pA. (d) Measured differential conductance vs voltage bias at three different bath temperatures for sample E. Arrows indicate the positions of $\Delta_1-\Delta_2$ (brown),  $\Delta_1+\Delta_2$ (blue) and 2$\Delta_1$ (green).}

\label{fig:sample}
\end{figure}

In this paper, we demonstrate experimentally that the hysteresis of the SQUIPT is removed and the responsivity of the device is enhanced this way. We further optimize the responsivity by shortening the weak link, decreasing the resistance of the tunnel junction and by making the latter narrower to act in good approximation as a local probe in the middle of the weak link. 

Figure 1 depicts the fabrication process starting from the shadow evaporation mask to the final device including a simplified measurement setup. The samples were fabricated by electron-beam lithography (EBL) onto an oxidized Si wafer using a bilayer resist which consists of a 900 nm thick copolymer layer and a 50 nm thick layer of PMMA ~\cite{Dolan77}. The EBL step is followed by development in MIBK:IPA 1:3 solution for 20 seconds, rinsing in IPA and drying. The detailed metallization steps are shown in Fig. 1(a):  the metals are deposited by electron-gun evaporation: first, 15 nm of Al at an angle $\theta=31^\circ$ is deposited and oxidized for 2 min with oxygen pressure of 1 mbar to form the tunnel barrier of the normal metal-insulator-superconductor (NIS) probe. The NIS junction resistance is about 100 k$\Omega$. Next, approximately 20 nm of copper at $\theta=-31^\circ$ is evaporated to complete the NIS ~\cite{Nahum94,Leivo96,Clark04} junction and to form the normal metal island. Finally, the sample is placed at an angle $\theta=10^\circ$ and the superconducting Al loop with 100 nm thickness is deposited to form a clean contact with the copper island. This step of deposition of a superconducting layer that is five times thicker compared to the normal metal forms the loop with reduced inductance and ensures that the inverse proximity effect is suppressed effectively. Figure 1(b) shows a scanning electron microscope (SEM) image of one of the fabricated SQUIPT devices with an enlarged view of the zone around the weak link with the attached tunnel probe in Fig. 1(c). 
The measurements of samples were performed in a $^3$He -$^4$He dilution refrigerator~\cite{Pekola94} at the temperature of 50 mK using preamplifiers at room temperature. 

Each SQUIPT is characterized by its physical dimensions, supercurrent $I_S$, which is  measured through the tunnel junction close to zero voltage bias and appearing in a normal conductor with superconducting proximity effect ~\cite{Belzing99}, and resistance $R$. Table I shows the values of these parameters for the measured samples. The NIS junction width $w$ and its normal resistance $R$ are also given. 
\begin{figure}[htb] 
\includegraphics[width =1\columnwidth]{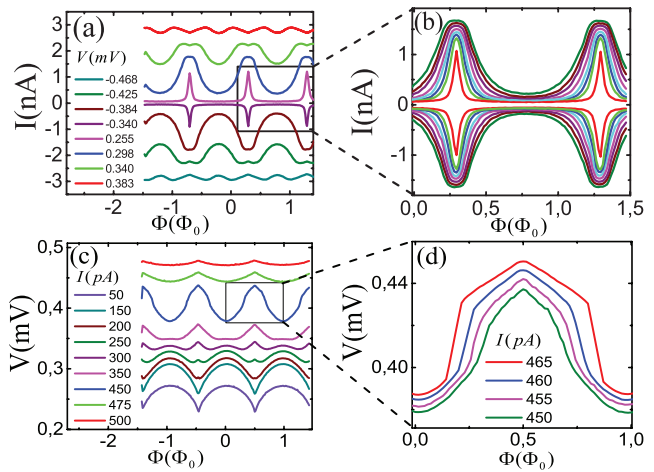}
\caption{\figta   Current modulation $I(\Phi)$ of the NIS junction at different values of bias voltage applied to it at $T_{bath}=~50$~ mK. (b) A zoomed view for several values of $V$ in the range $-340\mu$V$<V<298\mu$V. (c) Measured flux-to-voltage $V(\Phi)$ curves at different magnitudes of current through the junction and (d) a zoomed image at some values of current in the range 450 - 465 pA. The curves are not symmetric around $\Phi=0$ due to a constant offset flux bias.}

\label{fig:sample}
\end{figure}

A significant parameter of the normal-metal wire is its length. For a strong proximity effect in our samples, $L$ should be of the order of the superconducting coherence length $\xi_0$, $L=\alpha \xi_0$ where $\xi_0=(\hbar D/ \Delta_1)^{1/2}$ and $\Delta_1$  is the superconducting gap of aluminum leads. As representative parameters, we set $D= 0.01$~m$^2$S$^{-1}$ and $\Delta_1=220~ \mu$eV for all samples and then the magnitude of $\alpha$ is $\sim1.37, 1.45, 1.5$ and 1.62 for samples A to F. 


If the self inductance of the superconducting loop of the SQUIPT is negligible, the phase difference across the normal metal is determined by $\varphi= 2\pi \Phi / \Phi_{\circ}$  where $\Phi$ is the external magnetic flux through the loop.~\cite{Giazotto10,Meschke11}
Figure 2(a) represents experimental current-voltage (I-V) characteristics of sample A. In a diffusive regime of a SNS junction~\cite{Dubos08,Baselmans99}, the mini-gap in the normal-metal is of the order of Thouless energy $E_{th}= D/L^2$.~\cite{Pannetier0,Baselmans99} Based on the data with different values of magnetic flux in Fig. 2(b), the magnitude of the minigap $\Delta_2$ opened in the normal metal is minimized for $\varphi=\pi$  $(\Phi= \Phi_0/2)$  and maximized for $\varphi=0$ $(\Phi= 0)$.~\cite{Sueur08,Zhou98} In Fig. 2(c), the maximum supercurrent $I_S$ at $\Phi=0$  is approximately  48 pA  for sample A around zero bias voltage at $T = 50$ mK. The magnitude of this current is enhanced due to the stronger proximity effect in the present design and it is further increased with decreasing tunnel junction resistance as listed in Table \ref{tab:hresult}. It is negligible for the operation of the device as it stays well below the typical current bias values at the optimum working point of the SQUIPT.

As a consequence of mini-gap variations, the voltage and current modulations $V(\Phi)$, $I(\Phi)$, can be investigated at different values of current $I$ and voltage $V$ applied to the tunnel junction. Figure 3(a) illustrates current modulation of sample A for a number of bias voltages in the range from -0.468 mV to 0.383 mV. Figure 3(b) shows a detailed view of such curves at several values of bias voltages between -384 and $298~\mu$eV with a step size of approximately $4~ \mu$V. The corresponding flux-to-voltage $V(\Phi)$ characteristics are shown in Fig. 3(c) and Fig. 3(d) for several values of the bias current $I$. From Fig. 3, it is obvious that the hysteresis is absent at low temperatures in contrast to earlier work.~\cite{Meschke11}

The flux-to-current transfer function, $\partial I/\partial \Phi$, has been obtained by numerical differentiation of the $I(\Phi)$ characteristics. An example is shown in Fig. 4. The blue curve corresponds to the transfer function at $V=~ 0.251$ mV for which the device responsivity reaches $|\partial I/\partial \Phi|_{max}$$\cong ~23$ nA$/ \Phi_0$  at $T_{bath}=50$ mK. 

This value is about one order of magnitude higher than what has been reached earlier, because we implement here the way to overcome the unwanted hysteresis effect. This is the main result of the present work. Typical low noise room temperature current pre-amplifiers reach noise levels for high impedance sources of 5 $ \rm{fA}/ \sqrt{Hz}$ wich allows with the here reported responsivity a flux resolution of $\simeq 0.2 \times 10^{-6}  \Phi_{\circ}$.  This improvement is characterized also by the flux-to-voltage transfer function with maximum responsivity $|\partial V/\partial \Phi|_{max}$$\cong 1.7$~ mV$/ \Phi_0$  at base temperature. We note that the quality of the clean interface between the copper and aluminum influences the final performance of the practical devices as listed in Table \ref{tab:hresult}.

\begin{figure}[htb] 
\includegraphics[width = 1\columnwidth]{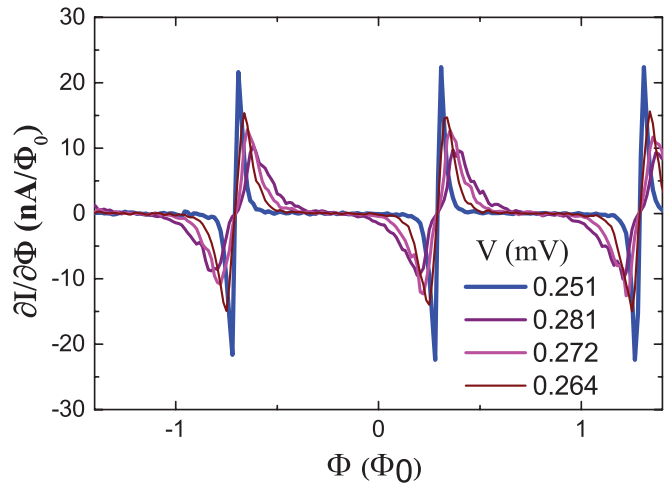} 
\caption{Current responsivity $\partial I/\partial \Phi$ characteristics as a function of the external magnetic flux through the superconducting loop of sample A at $T_{bath} = 50$ mK at four different values of applied bias voltage close to the optimum working point.}

\label{fig:sample}
\end{figure}

\begin{table}[h]
\caption{ Parameters of different samples measured at $T_{bath}=50$ mK. Here $d$, $L$, $w$, are the width and length of the copper island, and the width of the probe respectively. Al superconducting loop and Cu are 100 nm and $a=$ 20 nm thick respectively. The resistance of the NIS junction, $R$, was measured at low temperature and the maximum supercurrent of the probe is given by $I_S$. The maximum current and voltage responsivity as a function of magnetic flux are shown as $|\partial I/\partial \Phi|_{max}$ and $|\partial V/\partial \Phi|_{max}$.} 
\centering
\begin{tabular}{c ccccccc}
\hline\hline
sample& $L$& $d $ &  $w$   &     $R$          & $I_S$    & $|\frac{\partial I} {\partial \Phi}|_{max}$ & $|\frac{ \partial V}{\partial \Phi}|_{max}$ \\
           &(nm)&  (nm)   &         (nm)          & (k$\Omega)$ &  (pA)  &      $(\frac{nA} {\Phi_0})$             & $(\frac{mV} {\Phi_0})$  \\

\hline           
A & 237 &45& 70 &104& 48& 23&1.7\\
B & 250 &50 &80 & 243& 7& 8.4& 1.5\\
C & 250 &55& 80 &178& 5& 6.5&0.47\\
D & 250 &65&80 &145& 12& 7&0.45\\
E & 275 &66&107 &137& 10& 4&0.55\\
F & 280 &50& 90 &188& 9& 2&1\\
\hline\hline
\end{tabular}
\label{tab:hresult}
\end{table}

To further enhance the magnetic field responsivity of these devices, the area of the Al superconducting loop can be increased.~\cite{Gollop03} 
In our present device, we increased the Josephson inductance by factor of five as compared to the earlier work. 
With advanced lithography, the cross section of the weak link can be further reduced by about one order of magnitude: good quality N wires as narrow as $d=8$ nm can potentially be achieved by electron beam lithography and low temperature deposition.~\cite{Zgirski05,Wu06}  Hence, the superconducting loop size could be increased by the factor of 10, simply by increasing the loop diameter. As a consequence, enhancement of magnetic flux responsivity is feasible in these devices by the same factor.

The minigap in our devices reaches a magnitude of approximately 0.6 to 0.7 of the full superconducting gap $\Delta$. A further reduction of the weak link length to about 110 nm would already yield a minigap size of 0.97$\Delta$,~\cite{Cuevas06} corresponding to an improvement of the sensitivity of less than 30\%. On the other hand, replacing the superconductors with one with increased gap like vanadium~\cite{Ronzani} or niobium~\cite{Giazotto10} would allow almost an order of magnitude enhancement of the response function.


In conclusion, we have demonstrated experimentally that tuning the Josephson inductance of the weak link removes the unwanted hysteresis of such a device without affecting the response function. The latter is mainly defined by the weak link length and can be optimized independently. These findings allow one to design ultra-sensitive magnetometers which could lead to advancements in the field of nanoscale magnetometers at low temperatures.
 \newline

The work has been supported partially by the Iranian Ministry of Science, Research and Technology and by the Academy of Finland Centre of Excellence program (project number 250280). We acknowledge Micronova Nanofabrication Centre of Aalto University for providing the processing facilities.


\begin{thebibliography}{99}


\bibitem{Gennes66} P. G. de Gennes, $Superconductivity~  of~  Metals~  and ~ Alloys$ (W. A. Benjamin, New York, 1966).
 \bibitem{Tinkham96}M.Tinkham, $Introduction~ to ~Superconductivity$, 2nd Ed. (McGraw-Hill, New York, 1996).
 \bibitem{Gennes64} P. G. de Gennes, Rev. Mod. Phys. {\bf36}, 225 (1964).
 \bibitem{Buzdin05}A. I. Buzdin, Rev. Mod. Phys. {\bf77}, 935 (2005).
 \bibitem{Belzing99}W. Belzig {\it et al.}, Superlattices and Microstructures. {\bf 25}, 1251-1288 (1999).
\bibitem{Pannetier0}B. Pannetier and H. Courtois, J. Low Temp. Phys. {\bf 118}, 599 (2000).
\bibitem{Zhou98} F. Zhou, P. Charlat, B. Spivak, and B. Pannetier,  J. Low Temp. Phys. {\bf 110}, 841-843 (1998).
\bibitem{Giazotto10} F. Giazotto, J. T. Peltonen, M. Meschke, and J. P. Pekola, Nat. Phys.  {\bf 6}, 254 (2010).
\bibitem{Meschke11} M. Meschke, J. T. Peltonen, J. P. Pekola, and F. Giazotto, Phys. Rev. Lett. B.  {\bf 84}, 214514 (2011).

\bibitem{Foley09}C. P. Foley and H. Hilgenkamp, Supercond. Sci. Technol.  {\bf22}, 1-5 (2009).
\bibitem{Hao05}L. Hao {\it et al.}, IEEE Trans. Appl. Supercond. {\bf 15}, 514-517 (2005).
\bibitem{Hao07}L. Hao {\it et al.}, IEEE Trans. Instrum. Meas. {\bf 56}, 392-395 (2005).



\bibitem{Andreev 64} A. F. Andreev, Sov. Phys. JETP. {\bf 19}, 1228 (1964). 
\bibitem{Petrashov94}V. T. Petrashov, V. N. Antonov, P. Delsing, and T. Claeson, JETP Lett. {\bf 59}, 551 (1994).
\bibitem{Petrashov95}V. T. Petrashov, V. N. Antonov, P. Delsing, and T. Claeson, Phys. Rev. Lett. {\bf 74}, 5268 (1995).
\bibitem{Belzing02}W. Belzig, R. Shaikhaidarov, V. V. Petrashov, Yu. V. Nazarov, Phys. Rev. B. {\bf 66}, 220505 (2002).
\bibitem{Alidoust13} M. Alidoust, K. Halterman, and J.Linder, Phys. Rev. B. {\bf 88}, 075435 (2013).

\bibitem{Galaktionov12} A.V. Galaktionov, A.D. Zaikin, and L.S. Kuzmin, Phys. Rev. B. {\bf 85}, 224523 (2012).
\bibitem{Galaktionov13} A.V. Galaktionov, and A.D. Zaikin, Phys. Rev. B. {\bf 88}, 104513 (2013).

\bibitem{You11 } J. Q. You, and F. Nori, Nature. {\bf 474}, 589 (2011).

\bibitem{Giazotto08} F. Giazotto {\it et al.}, Appl. Phys. Lett. {\bf 92}, 162507 (2008).  
\bibitem{Blum04}Y. Blum, A. Tsukernik, M. Karpovsk, and J. M. Martinis, Phys. Rev. B. {\bf 70}, 214501 (2004).
\bibitem{Dubos08}P. Dubos, {\it et al.},  Phys. Rev. B.  {\bf 63}, 064502 (2008).
\bibitem{Heikkila02}T.T. Heikkila, J. Sarkka, and F. K. Wilhelm, Phys. Rev. B. {\bf 66}, 184513 (2002).
\bibitem{Dolan77} G. J. Dolan, Appl. Phys. Lett. {\bf 31}, 337 (1971).
\bibitem{Nahum94}M. Nahum, T. M. Eiles, and J. M. Martinis, Appl. Phys. Lett. {\bf 65}, 3123 (1994).
\bibitem{Leivo96}M. M. Leivo, J. P. Pekola, and D. V. Averin, Appl. Phys. Lett. {\bf 68}, 1996 (1996).
\bibitem{Clark04}A. M. Clark {\it et al.}, Appl. Phys. Lett. {\bf 84}, 625 (2004).

\bibitem{Pekola94} J. P. Pekola and J. P. Kauppinen, Cryogenics.  {\bf 34}, 843 (1994).
\bibitem{Baselmans99} J. J. A. Baselmans {\it et al.},  Nature.  {\bf 397}, 43 (1999).
\bibitem{Sueur08}H. le Sueur, P. Joyez, H. Pothier, C. Urbina, and D. Esteve, Phys. Rev. Lett. {\bf 100}, 197002 (2008).
\bibitem{Gollop03}J. Gallop, Supercond. Sci. Technol. {\bf 16}, 1576 (2003).
\bibitem{Zgirski05}M. Zgirski {\it et al.}, Nano. Lett.{\bf 5}, 1029-1033 (2005).
\bibitem{Wu06}X. S. Wu, {\it et al.}, PRL. {\bf 96}, 127002 (2006). 
\bibitem{Cuevas06}J. C. Cuevas,  J. Hammer, J. Kopu, J. K. Viljas, and M. Eschrig, Phys. Rev. B.  {\bf 73}, 184505 (2006).

\bibitem{Ronzani}A. Ronzani,  M. Baillergeau, C. Altimiras, and F. Giazotto, Appl. Phys. Lett. {\bf 103}, 052603 (2013).

\end{thebibliography}
\end{document}